\documentclass[twocolumn,american]{article}
\usepackage[T1]{fontenc}
\usepackage[latin9]{inputenc}
\usepackage{color}
\usepackage{babel}
\usepackage{amsmath}
\usepackage{amssymb}
\usepackage{graphicx}
\usepackage{geometry}
\usepackage{wasysym}
\usepackage[pdfusetitle,
 bookmarks=true,bookmarksnumbered=false,bookmarksopen=false,
 breaklinks=false,pdfborder={0 0 1},backref=false,colorlinks=false]
 {hyperref}



\begin{document}
\title{The connection between electromagnetic stress tensors and the dielectric
constant in a medium}
\author{R. Dengler \thanks{ORCID: 0000-0001-6706-8550}\\
Munich, Germany}
\maketitle
\begin{abstract}
We perform explicit calculations for microscopic models to confirm
that a single, unique electromagnetic stress tensor exists in a dielectric
medium: the microscopic tensor $\sigma^{\#}$. We demonstrate that
the conventional macroscopic stress tensor of continuum electrodynamics\textemdash which
contains the derivative of the dielectric coefficient\textemdash is
merely a representation of $\sigma^{\#}{}'s$ spatial average. This
result establishes that, contrary to claims in the literature, both
this averaged electromagnetic stress and the hydrodynamic pressure
always possess a unique physical meaning.
\end{abstract}

\section{Stress tensor in a medium}

We start with a short review of the commonly accepted formal expression
for the electromagnetic stress tensor in a liquid. We restrict the
discussion to non-magnetizable media. The magnetic part of the stress
tensor thus always agrees with its vacuum counterpart. 

\subsection{Conventional stress tensor}

The commonly accepted expression for the electromagnetic stress tensor
of a non-magnetizable liquid reads (\cite{LL_FT1971},\cite{Jackson1975})
\begin{align}
\sigma_{ij}^{\mathrm{liq}}\left(\boldsymbol{E},\boldsymbol{B}\right) & =\tfrac{\epsilon_{0}}{2}\left[\delta_{ij}\left(\left(\epsilon_{r}-\rho\tfrac{\partial\epsilon_{r}}{\partial\rho}\right)\boldsymbol{E}^{2}+\boldsymbol{B}^{2}\right)\right.\label{eq:sigma_LL}\\
 & \qquad\left.-2\epsilon_{r}E_{i}E_{i}-2B_{i}B_{j}\right].\nonumber 
\end{align}
The symbols $\boldsymbol{E}$ and $\boldsymbol{B}$ denote the macroscopic
(average) fields, $\epsilon_{r}=\epsilon/\epsilon_{0}$ is the relative
dielectric constant of the liquid, and $\rho$ its mass density. The
derivative $\partial\epsilon_{r}/\partial\rho$ is to be taken at
constant temperature. The sign of the electromagnetic stress tensor
is a matter of convention. The sign used in (\ref{eq:sigma_LL}) has
the advantage that the diagonal components can be interpreted as a
pressure. The expression (\ref{eq:sigma_LL}) is derived in an elementary
way in the next section. The practical value of formula (\ref{eq:sigma_LL})
is limited, however, because the derivative $\partial\epsilon_{r}/\partial\rho$
is usually unknown.

The main objective of this work is to relate quantities of macroscopic
(average) electrodynamics to microscopic quantities, which we denote
by a superscript '\#'. The microscopic stress tensor $\sigma_{ij}^{\#}\left(\boldsymbol{E}^{\#},\boldsymbol{B}^{\#}\right)$
has the same form as (\ref{eq:sigma_LL}), with $\boldsymbol{E}$
and $\boldsymbol{B}$ replaced by the microscopic fields $\boldsymbol{E}^{\#}$
and $\boldsymbol{B}^{\#}$ and with $\epsilon_{r}$ replaced with
unity. This also implies $\partial\epsilon_{r}/\partial\rho=0$. The
microscopic stress tensor has an immediate physical meaning. The volume
force
\[
f_{i}^{\#}=-\sum_{j}\nabla_{j}\sigma_{ij}^{\#}\left(\boldsymbol{E}^{\#},\boldsymbol{B}^{\#}\right)=\rho_{e}^{\#}E_{i}^{\#}+\left(\boldsymbol{j}^{\#}\times\boldsymbol{B}^{\#}\right)_{i}
\]
is the sum of the Coulomb and Lorentz force densities acting on the
microscopic electric charge distribution. The force $\int_{V}\mathrm{d}^{d}xf_{i}^{\#}$
acting on matter within a macroscopic volume $V$ generally depends
on the often irregular distribution of the atoms near the surface
of the volume. The average, however, is expected to agree with that
implied by the macroscopic stress tensor~(\ref{eq:sigma_LL}). 

\subsection{Planar capacitor filled with a liquid}

We assume that a capacitor with area $A=w\ell$ and distance of the
plates $h$ completely encloses a liquid or a vacuum in a volume $V=Ah$,
and that the plates or walls can move. For constant charge $Q$ the
capacitor has the energy
\begin{equation}
W=\tfrac{h}{2\epsilon A}Q^{2}=\tfrac{\epsilon}{2}E^{2}V.\label{eq:W_capac}
\end{equation}
We take the height $h$ as $x_{2}$ direction and for clarity sometimes
write $\ell_{1}=w$, $\ell_{2}=h$ and $\ell_{3}=\ell$ in the results.
It is convenient to assume that the lateral walls are (hypothetical)
magnetic superconductors. Such walls are purely passive objects, which
allow a parallel electric field, but expel any electric field, and
thus eliminate fringe effects. 

The dielectric coefficient $\epsilon=\epsilon\left(\rho\right)$ is
a function of the mass density $\rho.$ In the case of a vacuum $\sigma_{22}^{\mathrm{vac}}=-\tfrac{\epsilon_{0}}{2}E_{2}^{2}$,
and the plates attract each other, with the force $-\tfrac{\epsilon_{0}}{2}E_{2}^{2}A$.
The stress tensor $\sigma_{11}^{\mathrm{vac}}=\tfrac{\epsilon_{0}}{2}E_{2}^{2}$
in the direction perpendicular to the electric field implies that
the lateral walls are pushed outward with the force $\tfrac{\epsilon_{0}}{2}whE_{2}^{2}$.

\subsubsection*{Force on lateral walls}

To simplify the discussion we ignore the hydrodynamic pressure. We
assume that the distance $w=\ell_{1}$ of the pair of lateral walls
in $x_{1}$-direction is variable. The stress tensor component in
$x_{1}$-direction follows from (\ref{eq:W_capac}) with the help
of $\partial\rho/\partial\ell_{1}=-\rho/\ell_{1}$ and $Q/\left(\ell_{1}\ell_{3}\right)=\epsilon E_{2}$
as 
\begin{equation}
\sigma_{11}^{\mathrm{liq}}\left(E_{2}\right)=-\tfrac{1}{\ell_{2}\ell_{3}}\left(\tfrac{\partial W}{\partial\ell_{1}}\right)_{Q}=\tfrac{1}{2}\left(\epsilon-\rho\tfrac{\partial\epsilon}{\partial\rho}\right)E_{2}^{2}.\label{eq:sigma11_capac}
\end{equation}
This agrees with Eq.~(\ref{eq:sigma_LL}) for $B=0$. A calculation
with constant voltage leads to the same result. One could connect
the capacitor to a voltage source with constant voltage $U$. The
charge then is $UC$, and the total energy $W$ now includes the energy
$-QU$ of the voltage source. The derivation of (\ref{eq:sigma22_capac})
of course is equivalent to that in (\cite{LL_FT1971}) based on virtual
deformations, but it uses a precisely defined physical scenario. 

\subsubsection*{Force on plates (in field direction)}

The change of the energy~(\ref{eq:W_capac}) with the height $h=\ell_{2}$
implies the stress tensor component
\begin{equation}
\sigma_{22}^{\mathrm{liq}}\left(E_{2}\right)=-\tfrac{1}{A}\left(\tfrac{\partial W}{\partial h}\right)_{Q}=-\tfrac{1}{2}\left(\epsilon+\rho\tfrac{\partial\epsilon}{\partial\rho}\right)E_{2}^{2}\label{eq:sigma22_capac}
\end{equation}
in the direction of the electric field. Here we have used $\partial\rho/\partial h=-\rho/h$
and $Q/A=\epsilon E$. The result agrees with (\ref{eq:sigma_LL})
in the case $B=0$.

\subsubsection*{Constant mass density}

\label{subsec:rho_const}According to the above results the plates
are pulled inward and the lateral walls are pushed outward. In the
derivation it was assumed that only one boundary moves at a time,
which changes the mass density. We have ignored any mechanical stress,
but the dependence of the dielectric coefficient on $\rho$ was taken
into account. 

In the case of liquids there are situations where the mass density
remains constant. An example is a capacitor without lateral walls
or a small hole anywhere immersed in a liquid. The system energy now
is $W$ from Eq.~(\ref{eq:W_capac}) with constant dielectric coefficient
$\epsilon$. If the length is changed to $x_{1}'=\ell+\delta\ell$
then the energy changes to $W'=W-W\delta\ell/\ell$. This gives the
first line of 
\begin{align}
\sigma_{11|\rho}^{\mathrm{liq,tot}}\left(E_{2}\right) & =-\tfrac{1}{hw}\left(\tfrac{\delta W}{\delta\ell}\right)_{Q,\rho}=\tfrac{\epsilon}{2}E_{2}^{2},\label{eq:sigma11_rho}\\
\sigma_{22|\rho}^{\mathrm{liq,tot}}\left(E_{2}\right) & =-\tfrac{1}{\ell w}\left(\tfrac{\delta W}{\delta h}\right)_{Q,\rho}=-\tfrac{\epsilon}{2}E_{2}^{2}.\label{eq:sigma22_rho}
\end{align}
The second line follows from the dependence of the energy (\ref{eq:W_capac})
on the height $x_{2}=h$ for constant $\epsilon$. We have used the
label ``tot'' for ``total'' because this now is an open system
and the energy balance gives no information about the nature of the
forces. However, the fact that the capacitor has a hole and that matter
can flow in and out does not change the purely electrostatic contribution
(\ref{eq:sigma11_capac}, \ref{eq:sigma22_capac}) to the force acting
on the boundaries. We thus conclude that (\ref{eq:sigma11_rho},\ref{eq:sigma22_rho})
is the sum of (\ref{eq:sigma11_capac}) or (\ref{eq:sigma22_capac})
and a hydrodynamic pressure
\begin{equation}
p=\tfrac{\epsilon_{0}}{2}\rho\tfrac{\partial\epsilon_{r}}{\partial\rho}\boldsymbol{E}^{2}.\label{eq:pressure}
\end{equation}
A slight complication arises from the fact that the pressure (\ref{eq:pressure})
compresses the liquid somewhat and thus modifies $\epsilon$ nevertheless.
The ensuing $\delta\epsilon=O\left(E^{2}\right)$, however, is a second
order effect and can be ignored.

It is a source confusion that the pressure (\ref{eq:pressure}) contributes
to $\sigma_{ii|\rho}^{\mathrm{liq,tot}}$ but does not show up in
the expressions (\ref{eq:sigma11_rho}) and (\ref{eq:sigma22_rho}),
while it formally occurs in the expressions (\ref{eq:sigma11_capac})
and ( \ref{eq:sigma22_capac}) for $\sigma^{\mathrm{liq}},$ where
no pressure is in play.

The stress (\ref{eq:sigma22_rho}) has been confirmed in numerous
experiments. We do not agree with the theoretical work \cite{Zanch_1994}
where $\left(\epsilon/2\right)E_{2}^{2}$ is interpreted as Coulomb
force without a contribution from pressure. The point we want to make
is that the electromagnetic stress tensor is an intensive quantity
with a unique meaning as the average of the electromagnetic vacuum
stress tensor. This is verified below with exact solutions for microscopic
models.

\textcolor{magenta}{}

\subsection{Planar capacitor containing a solid}

The above derivation of the electrostatic stress tensor (\ref{eq:sigma11_capac},
\ref{eq:sigma22_capac}) of a liquid can be generalized to solids
if the principle axis of the dielectric tensor are parallel to the
borders of the capacitor, and if $\epsilon$ is replaced with the
component $\epsilon_{22}$ of the dielectric tensor in field direction.
Here we only consider amorphous media or cubic crystals with originally
$\epsilon_{ij}=\delta_{ij}\epsilon$. An uniaxially compressed liquid
remains isotropic and thus $\epsilon=\epsilon\left(\rho\right)$.
This is not true for solids where the compression changes the aspect
ratio of the unit cells. In general thus $\ell_{2}\partial\epsilon_{22}/\partial\ell_{2}\neq\ell_{1}\partial\epsilon_{22}/\partial\ell_{1}$,
and the energy balance leads to
\begin{align}
\sigma_{11}^{\mathrm{solid}}\left(E_{2}\right) & =\tfrac{1}{2}\left(\epsilon_{22}+\ell_{1}\tfrac{\partial\epsilon_{22}}{\partial\ell_{1}}\right)E_{2}^{2},\label{eq:sigma11_solid}\\
\sigma_{22}^{\mathrm{solid}}\left(E_{2}\right) & =-\tfrac{1}{2}\left(\epsilon_{22}-\ell_{2}\tfrac{\partial\epsilon_{22}}{\partial\ell_{2}}\right)E_{2}^{2}.\label{eq:sigma22_solid}
\end{align}
The derivatives describe the onset of anisotropy under uniaxial deformations
in directions perpendicular and parallel to the electric field.

For an undeformed cubic or amorphous solid the expressions (\ref{eq:sigma11_solid},
\ref{eq:sigma22_solid}), like Eq.~(\ref{eq:sigma_LL}), are the
electrostatic stress tensor for a given undeformed medium. In principle
the values can also be calculated from a microscopic model. The derivatives
in (\ref{eq:sigma11_solid}) and (\ref{eq:sigma22_solid}) as well
as in (\ref{eq:sigma_LL}) thus are misleading.

\section{Uniaxial microscopic model}

We now examine two microscopic models allowing to calculate the vacuum
stress tensor in planes without matter in the medium. The average
vacuum stress tensor as expected agrees with the more phenomenological
expression~(\ref{eq:sigma_LL}). 

\begin{figure}

\begin{centering}
\includegraphics[scale=0.5]{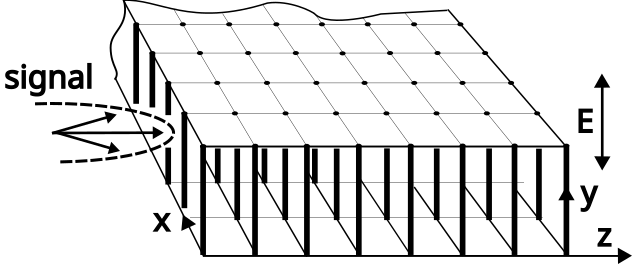}\caption{\label{fig:RodModel}A dielectric medium consisting of polarizable
rods. Microscopic and macroscopic electric field are identical.}
\par\end{centering}
\end{figure}
A rather simple model consists of a medium built of thin polarizable
rods oriented in $x_{2}$-direction on a square lattice in the $x_{1}$-$x_{3}$-plane,
see Fig.~\ref{fig:RodModel}. It is assumed that the electric field
points in $x_{2}$-direction. No polarization charge arises in the
medium. The electric field thus is constant and no Clausius-Mossotti
relation is required. Only the $\epsilon_{22}$ component of the dielectric
tensor plays a role. In this system macroscopic and microscopic fields
agree. If the system has length $\ell_{1}$ in $x_{1}$ direction
(perpendicular to the rods) then $\epsilon_{22}=\epsilon_{0}\left(1+\gamma_{a}/\ell_{1}\right)$
where $\gamma_{a}$ is a constant. Eq.~(\ref{eq:sigma11_solid})
then leads to
\[
\sigma_{11}^{\mathrm{rod}}\left(\boldsymbol{E}\right)=\tfrac{1}{2}\left(\epsilon_{22}+\ell_{1}\tfrac{\partial\epsilon_{22}}{\partial\ell_{1}}\right)E_{2}^{2}=\tfrac{\epsilon_{0}}{2}E_{2}^{2}.
\]
The stress tensor~(\ref{eq:sigma11_solid}) thus correctly reproduces
the vacuum stress tensor in the planes $x_{1}=\mathrm{const}$ parallel
to the rods and the electric field. Evidently the model also allows
exact solutions for signal propagation in directions perpendicular
to the $x_{2}$ axis. 

\section{Point dipoles on an orthorhombic lattice}

As a more realistic model we examine in detail an orthorhombic crystal.
The medium consists of polarizable atoms built from two interpenetrating
spherical charge distributions $\pm Q_{a}F\left(\boldsymbol{x}\right)$
with a charge $\pm Q_{a}$ on a rectangular lattice with lattice constants
$a_{i}$ and cell volume $\Omega=a_{1}a_{2}a_{3}$. We assume that
the radial scale $R$ of $F\left(\boldsymbol{x}\right)$ satisfies
$R\ll\min\left(a_{1},a_{2},a_{3}\right)$. Three simple charge distributions,
together with the Fourier transform $\tilde{F}$ are
\begin{align}
\begin{array}{c|c}
F\left(\boldsymbol{x}\right) & \tilde{F}\left(qR\equiv s\right)\\
\hline e^{-\boldsymbol{x^{2}}/R^{2}}/\left(\sqrt{\pi}R\right)^{3} & e^{-s^{2}/4}\\
e^{-\left|\boldsymbol{x}\right|/R}/\left(8\pi R^{3}\right) & 1/\left(1+s^{2}\right)^{2}\\
\theta\left(R^{2}-\boldsymbol{x}^{2}\right)/\left(\tfrac{4\pi}{3}R^{3}\right) & 3\left(\sin\left(s\right)-s\cos\left(s\right)\right)/s^{3}.
\end{array}\label{eq:F_Funct}
\end{align}
If $Q_{a}$ is large then the charge density of the atom at the origin
in a local field $\boldsymbol{E}^{\mathrm{loc}}$ is
\begin{align}
\rho_{\mathrm{dip}}^{\#\left(0\right)}\left(\boldsymbol{x}\right) & =-\boldsymbol{p}\nabla F\left(\boldsymbol{x}\right)=-\epsilon_{0}\gamma\Omega\boldsymbol{E}^{\mathrm{loc}}\cdot\nabla F\left(\boldsymbol{x}\right),\label{eq:rho_dip0}
\end{align}
where we have used
\begin{equation}
\boldsymbol{p}=\epsilon_{0}\gamma\Omega\boldsymbol{E}^{\mathrm{loc}}.\label{eq:gamma_def}
\end{equation}
The polarizability of an atom is $\gamma_{a}=\gamma\Omega$. The
charge density of the medium can be written as a Fourier series,
\begin{align*}
\rho^{\#}\left(\boldsymbol{x}\right) & =-i\epsilon_{0}\gamma\boldsymbol{E}^{\mathrm{loc}}\cdot\sum_{q_{i}\in2\pi\mathbb{Z}/a_{i}\setminus0}e^{i\boldsymbol{qx}}\boldsymbol{q}\tilde{F}\left(qR\right),\\
\tilde{F}\left(qR\right) & =\int\mathrm{d}^{3}xe^{-i\boldsymbol{qx}}F\left(\boldsymbol{x}\right).
\end{align*}
We now assume that $\boldsymbol{E}^{\mathrm{loc}}$ and the average
(macroscopic) field $\boldsymbol{E}$ point in $x_{2}$ direction.
The Fourier series of the electric field follows as 
\begin{align}
\boldsymbol{E}^{\#}\left(\boldsymbol{x}\right) & =\boldsymbol{E}+\tilde{\boldsymbol{E}}\left(\boldsymbol{x}\right),\label{eq:E=000023_ortho}\\
\tilde{\boldsymbol{E}}\left(\boldsymbol{x}\right) & =-\gamma E_{2}^{\mathrm{loc}}\sum_{\boldsymbol{q}\in\left(2\pi\mathbb{Z}/a_{i}\right)\setminus0}e^{i\boldsymbol{qx}}\tfrac{\boldsymbol{q}q_{2}}{q^{2}}\tilde{F}\left(qR\right)\cdot\nonumber 
\end{align}
The usual definition $p_{2}/\Omega=\left(\epsilon_{22}-\epsilon_{0}\right)E_{2}$
of the dielectric constant in terms of polarization density and average
field $E_{2}$ leads to
\begin{equation}
\left(\epsilon_{22}-\epsilon_{0}\right)E_{2}=\epsilon_{0}\gamma E_{2}^{\mathrm{loc}}.\label{eq:eps_2_gamma}
\end{equation}
This allows to eliminate the microscopic quantity $\gamma E_{2}^{\mathrm{loc}}$
from Eq.~(\ref{eq:E=000023_ortho}). The microscopic model, however,
also allows to calculate the dielectric constant. This can be achieved
by writing down the field $\boldsymbol{E}^{\#}\left(\boldsymbol{x}=0\right)$
at the origin in a different way. The alternative expression is the
sum of $E_{2}^{\mathrm{loc}}$ and the field generated by the charge
distribution~(\ref{eq:rho_dip0}) of the atom at the origin at $\boldsymbol{x}=0$,
\begin{equation}
E_{i}^{\mathrm{dip}}=\tfrac{-1}{4\pi\epsilon_{0}}\int\mathrm{d}^{3}x\rho_{\mathrm{dip}}^{\#\left(0\right)}\left(\boldsymbol{x}\right)x_{i}/\left|\boldsymbol{x}\right|^{3}=-\tfrac{\gamma}{3}\Omega F\left(0\right)E_{2}^{\mathrm{loc}}\delta_{i,2}.\label{eq:E=0000230_E_loc}
\end{equation}
Equating now $E_{2}^{\mathrm{loc}}+E_{i}^{\mathrm{dip}}$ and Eq.~(\ref{eq:E=000023_ortho})
for $\boldsymbol{x}=0$ and eliminating $E_{2}$ with (\ref{eq:eps_2_gamma})
leads to

\begin{align}
\epsilon_{22}/\epsilon_{0} & =1+\gamma/\left(1-\gamma\Lambda\right)\equiv\epsilon_{r},\label{eq:eps2_from_sum}\\
\Lambda & =\tfrac{\Omega}{3}F\left(0\right)-\sum_{\boldsymbol{q}\in\left(2\pi\mathbb{Z}/a_{i}\right)\setminus0}\tfrac{q_{2}^{2}}{q^{2}}\tilde{F}\left(qR\right).\label{eq:Lambda_def}
\end{align}
These equations allow to calculate the dielectric constant of orthorhombic
crystals. The generalization more complicated crystals would not pose
any problems. Atoms with a Gaussian charge distribution $\tilde{F}\left(x\right)=e^{-x^{2}/4}$
lead to a rapidly convergent sum. In fact, Eq.~(\ref{eq:Lambda_def})
coincides with the reciprocal lattice part of the corresponding Ewald
sum for point dipoles (the radius $R$ plays the role of the convergence
parameter), the lattice part is of order $e^{-a^{2}/R^{2}}$ and negligible
for $R\ll a$. This confirms that the $R=0$ limit is equivalent to
point dipoles. 

The Ewald sum has been used in~\cite{LoWanYu2001} to examine the
dielectric constant of tetragonal crystals. The authors observe that
the Ewald sum reaches a plateau value for a special value $\eta$,
while in fact it is independent of $\eta$. This indeed is the case
if the $\eta^{3}/\sqrt{\pi}$ in Eq.~(8) in~\cite{LoWanYu2001}
is replaced with the correct $\left(\eta/\sqrt{\pi}\right)^{3}$.
The sum then rapidly converges for any deformation, as it is the case
with the sum in Eq.~(\ref{eq:Lambda_def}).

The result $\Lambda=1/3$ for a cubic crystal $a_{i}=a$ leads to
the usual Clausius-Mossotti formula for $\epsilon$. The alternative
charge distributions $F\left(\boldsymbol{x}\right)$ from Eq.~(\ref{eq:F_Funct})
lead to the same values in the $R=0$ limit. The convergence, however,
is much slower. We will use the formula~(\ref{eq:eps2_from_sum})
to verify the relations~(\ref{eq:sigma11_solid}) and~(\ref{eq:sigma22_solid})
between $\epsilon$ and the electromagnetic stress tensor.

\subsection{Stress tensor perpendicular to electric field}

We now use the electric field (\ref{eq:E=000023_ortho}) to calculate
the average electromagnetic stress tensor in the vacuum plane $x_{1}=a/2$
between the (point-like) dipoles in a cubic crystal. If $\boldsymbol{E}$
points in $x_{2}$-direction then the granular part of $\boldsymbol{E}^{\#}$
can be written as
\begin{align}
\tilde{E}_{i} & =\left(\epsilon_{r}-1\right)E_{2}\partial_{i}\partial_{2}\tilde{S},\label{eq:E_tilde}\\
\tilde{S} & =\sum_{\boldsymbol{q}\in\left(2\pi\mathbb{Z}/a\right)^{3},q_{2}\neq0}e^{i\boldsymbol{qx}}/\boldsymbol{q}^{2}.\label{eq:S_sum}
\end{align}
Because of the derivative $\partial_{2}$ in $\tilde{E}$ the sum
in $\tilde{S}$ can be restricted to $q_{2}\neq0$ instead of $\boldsymbol{q}\neq0$.
The sum over $q_{1}$ in $\tilde{S}$ can then be performed in closed
form, for instance as a Matsubara sum. The result is 
\begin{align}
\tilde{S} & =\tfrac{a^{2}}{4}\sum_{q_{2}\neq0,q_{3}}e^{i\left(q_{2}x_{2}+q_{3}x_{3}\right)}\tfrac{\cosh\left(Q\left(x_{1}-\tfrac{a}{2}\right)\right)}{\tfrac{Qa}{2}\sinh\tfrac{Qa}{2}}.\label{eq:Matsubara sum}
\end{align}
This leads to
\[
\tilde{E}_{i\in\left\{ 2,3\right\} }\left(x_{1}=\tfrac{a}{2}\right)=-\left(\epsilon_{r}-1\right)E_{2}\sum_{q_{2}\neq0,q_{3}}e^{i\left(q_{2}x_{2}+q_{3}x_{3}\right)}\tfrac{a^{2}q_{i}q_{2}/4}{\tfrac{Qa}{2}\sinh\tfrac{Qa}{2}}.
\]
Contributions to $\sigma^{\#}$ of the form $E_{i}\tilde{E_{j}}$
vanish when averaged over an area $a^{2}$, and the only nontrivial
contributions are of the form $\tilde{E}\tilde{E}$. It follows 
\begin{align}
\left\langle \tilde{\sigma}_{11}\left(x_{1}=\tfrac{a}{2}\right)\right\rangle  & =\tfrac{\epsilon_{0}}{2}\left(\epsilon_{r}-1\right)^{2}E_{2}^{2}M,\nonumber \\
M & \overset{\checkmark}{=}\tfrac{1}{2}\sum_{\left\{ q_{2},q_{3}\right\} \neq0}\tfrac{\left(Qa/2\right)^{2}}{\sinh^{2}\tfrac{Qa}{2}}=\text{0.171599}\ldots,\nonumber \\
\left\langle \sigma_{11}^{\#,\mathrm{cubic}}\left(E_{2}\right)\right\rangle  & \overset{\checked}{=}\tfrac{\epsilon_{0}}{2}E_{2}^{2}\left(1+M\left(\epsilon_{r}-1\right)^{2}\right),\label{eq:sigma=00002311}
\end{align}
where $M$ is a rapidly convergent sum. 

\subsection{Stress tensor in direction of electric field}

To perform the sum over $q_{2}$ in (\ref{eq:S_sum}) we split the
sum into contributions with $Q^{2}=q_{1}^{2}+q_{3}^{2}=0$ and $Q^{2}\neq0$,
\begin{align}
\tilde{S} & =\sum_{q_{2}\neq0}e^{iq_{2}x_{2}}/q_{2}^{2}+\sum_{\boldsymbol{q}\in\left(2\pi\mathbb{Z}/a\right)^{3},\,Q\neq0}e^{i\boldsymbol{qx}}/\boldsymbol{q}^{2}\label{eq:STilde_q1q3}\\
 & =\tfrac{1}{2}\left(x_{2}^{2}-x_{2}a+\tfrac{a^{2}}{6}\right)+\tfrac{a^{2}}{4}\sum_{\left\{ q_{1},q_{3}\right\} \neq0}e^{i\left(q_{1}x_{1}+q_{3}x_{3}\right)}\tfrac{\cosh\left(Q\left(x_{2}-\tfrac{a}{2}\right)\right)}{\tfrac{Qa}{2}\sinh\tfrac{Qa}{2}}.\nonumber 
\end{align}
In the $Q\neq0$ part we have again used the formula~(\ref{eq:Matsubara sum}).
The first sum in (\ref{eq:STilde_q1q3}) is $\lim_{Q\rightarrow0}\left(f_{Q}\left(y\right)-1/Q^{2}\right)$
from the same equation. From $\partial_{2}\tilde{S}=0$ at $x_{2}=a/2$
it follows $\tilde{E}_{1}=\tilde{E}_{3}=0$, but there remains
\begin{align*}
\tilde{E}_{2}\left(x_{2}=\tfrac{a}{2}\right) & =\left(\epsilon_{r}-1\right)E_{2}\partial_{2}^{2}\tilde{S}\\
 & =\left(\epsilon_{r}-1\right)E_{2}\left(1+\sum_{\left\{ q_{1},q_{3}\right\} \neq0}e^{i\left(q_{1}x_{1}+q_{3}x_{3}\right)}\tfrac{Qa/2}{\sinh\tfrac{Qa}{2}}\right).
\end{align*}
The actual non-granular field in the $x_{2}=a/2$ plane thus is $\epsilon_{r}E_{2}$
instead of $E_{2}$. Averaging $\sigma_{22}^{\#}=-\tfrac{\epsilon_{0}}{2}E_{2}^{\#2}$
over an area $a^{2}$ leads to 
\begin{equation}
\left\langle \sigma_{22}^{\#,\mathrm{cubic}}\left(E_{2}\right)\right\rangle =-\tfrac{\epsilon_{0}}{2}E_{2}^{2}\left(\epsilon_{r}^{2}+2M\left(\epsilon_{r}-1\right)^{2}\right).\label{eq:sigma=00002322}
\end{equation}
The $\epsilon_{r}^{2}$ contribution is due to the constant in $\tilde{E}_{2}$.
This contribution was not taken into account in \cite{Deng2_2025}.

\subsection{Deformation of a cubic crystal}

For a cubic crystal we now have the phenomenological stress tensor
$\sigma_{11|22}^{\mathrm{solid}}$ from Eq.~(\ref{eq:sigma11_solid})
and Eq.~(\ref{eq:sigma22_solid}) and the microscopic stress tensor
$\sigma_{11|22}^{\#,\mathrm{cubic}}$ from (\ref{eq:sigma=00002322})
and (\ref{eq:sigma=00002322}). To be able to relate the values we
also need the variation of $\epsilon_{22}$ of a cubic crystal under
deformations $a\rightarrow a\xi_{i}$ with $\xi_{i}\cong1$. This
variation can be taken into account in the sum (\ref{eq:eps2_from_sum})
by replacing $q_{i}$ with $q_{i}/\xi_{i}$. Numerically one finds
for deformations in $x_{1}$ or $x_{2}$ direction
\begin{equation}
\epsilon_{r}=1+\tfrac{\gamma_{a}}{\Omega-\gamma_{a}\left[1/3+\left(M+\tfrac{1}{3}\right)\delta\xi_{1}-2\left(M+\tfrac{1}{3}\right)\delta\xi_{2}\right]}+\left(\delta\xi\right)^{2}\ldots,\label{eq:eps_r_ortho}
\end{equation}
where $\Omega=a^{3}\xi_{1}\xi_{2}\xi_{3}$ and $M\cong0.171599$ is
the constant from Eq.~(\ref{eq:sigma=00002311}).

We first consider an uniaxial deformation in the direction perpendicular
to the electric field. Equating (\ref{eq:sigma11_solid}) and (\ref{eq:sigma=00002311})
gives an expression for the variation of the dielectric constant,
\begin{equation}
\ell_{1}\partial\epsilon_{r}/\partial\ell_{1}=-\left(\epsilon_{r}-1-M\left(\epsilon_{r}-1\right)^{2}\right).\label{eq:eps_l1}
\end{equation}
The r.h.s. is negative for $\epsilon_{r}<1+2/M\cong12.65.$  The
expression~(\ref{eq:eps_l1}) agrees with the derivative $\left.\partial_{\xi_{1}}\epsilon_{r}\right|_{\xi=1}$
directly calculated from the lattice model value~(\ref{eq:eps_r_ortho}).

For an uniaxial deformation in the direction of the electric field
we can equate the expressions~(\ref{eq:sigma22_solid}) and~(\ref{eq:sigma=00002322}).
This similarly leads to
\begin{equation}
\ell_{2}\partial\epsilon_{r}/\partial\ell_{2}=-\left(\epsilon_{r}^{2}-\epsilon_{r}+2M\left(\epsilon_{r}-1\right)^{2}\right)<0.\label{eq:eps_l2}
\end{equation}
 The r.h.s. is negative for any $\epsilon_{r}>1$. The result (\ref{eq:eps_l2})
likewise agrees with the derivative $\left.\partial_{\xi_{2}}\epsilon_{r}\right|_{\xi=1}$
directly calculated from the lattice model value (\ref{eq:eps_r_ortho}).

The last variant is an isotropic deformation. A cubic crystal remains
cubic under such a deformation, and the Clausius-Mossotti relation
$\epsilon_{r}=1+\gamma_{a}/\left(\Omega-\gamma_{a}/3\right)$ from
the microscopic Eq.~(\ref{eq:eps2_from_sum}) is the exact solution.
Using $\rho\partial_{\rho}=-\Omega\partial_{\Omega}$ leads to
\[
\rho\partial_{\rho}\epsilon_{r}\overset{\checked}{=}\tfrac{\gamma}{\left(1-\gamma/3\right)^{2}}\overset{\checked}{=}\tfrac{1}{3}\left(\epsilon_{r}^{2}+\epsilon_{r}-2\right).
\]
The work done in the deformation now involves the weighted average
\[
\tfrac{2}{3}\sigma_{11}^{\mathrm{solid}}+\tfrac{1}{3}\sigma_{22}^{\mathrm{solid}}=\tfrac{\epsilon_{0}}{6}\left(2-\epsilon_{r}^{2}\right)E_{2}^{2},
\]
of Eq.~(\ref{eq:sigma11_solid}) and Eq.~(\ref{eq:sigma22_solid}),
and this again agrees with the weighted average of the microscopic
expressions (\ref{eq:sigma=00002311}) and (\ref{eq:sigma=00002322}).

\section{Conclusion}

We have demonstrated with exact solutions of microscopic models that
the microscopic electromagnetic stress tensor $\sigma^{\#}$ is the
central physical quantity.

From this perspective, the conventional macroscopic stress tensor
is merely a formula that reproduces the average of $\sigma^{\#}$
over a suitable area, such as a unit cell's boundary or another area
large enough for averaging. This could have been expected. It is also
clear that the conventional macroscopic stress tensor is of limited
value on itself as long as derivatives like $\rho\partial\epsilon/\partial\rho$
are unknown. However, together with $\sigma^{\#}$ the average stress
tensor always is defined.

One might object that the point dipole model studied above is something
special in that it contains planes devoid of matter. This is a misconception.
Real atoms are largely empty space. The quantum mechanical total stress
tensor (momentum flow) expressed in terms of an $N$-body wavefunction
or in second quantized form comprises a kinetic (quantum mechanical)
part and a Coulomb part. The Coulomb part is the microscopic electromagnetic
stress tensor.

It suggests itself to use the results derived above to describe propagating
electromagnetic signals with microscopic models. For point dipoles
this is essentially the discrete dipole approximation (DDA) \cite{Draine_1993}.
We have made an attempt in this direction in \cite{Deng2_2025}, but
some questions remain. For wavelengths much larger than the lattice
constant, most signal properties depend only on the amplitude, wavevector,
polarization direction and the dielectric constant. In essence the
propagation is a quasi-static phenomenon. Nevertheless confusion persists
regarding the mechanical and electromagnetic momentum flow (stress
tensor) in such signals in a medium. Two more recent reviews covering
this vast field are \cite{Kemp2011,Milonni_2010}. \bigskip{}

\bibliographystyle{habbrv}
\bibliography{Other}

\end{document}